\documentclass[twocolumn,prX,nofootinbib]{revtex4-1}
\pdfoutput=1

\usepackage{amsmath,amssymb,amsfonts,amsthm,array,eucal,mathrsfs,upgreek,stackrel,slashed,graphicx,tensor,color,enumerate}
\usepackage{hyperref}

\usepackage[usenames,dvipsnames,svgnames,table]{xcolor}

\DeclareMathAlphabet{\mathpzc}{OT1}{pzc}{m}{it}

\newcommand{\beqn}[1]{\begin{equation*}#1\end{equation*}}
\newcommand{\beql}[1]{\begin{equation}\label{#1}}
\newcommand{\eeq}{\end{equation}}
\newcommand{\ie}{\textit{i.e.}~}
\newcommand{\eg}{\textit{e.g.}~}
\newcommand{\ord}{\boldsymbol{\CMcal{O}}}
\newcommand{\dl}{\partial}
\newcommand{\Dl}{\nabla}
\newcommand{\vDl}{\vec{\nabla}}
\newcommand{\ed}{\mathrm{d}}
\newcommand{\dr}{\mathrm{d}r}
\newcommand{\ds}{\mathrm{d}s}
\newcommand{\dt}{\mathrm{d}t}
\newcommand{\dT}{\mathrm{d}T }
\newcommand{\dv}{\mathrm{d}v}
\newcommand{\dy}{\mathrm{d}y}
\newcommand{\met}{\mathsf{g}}
\newcommand{\CC}{\textsc{cc}}

\newcommand{\GHY}{\textsc{ghy}}
\newcommand{\lag}{\mathscr{L}}
\newcommand{\vAE}{\vec{\AE}}
\newcommand{\elL}{\ell_\textsc{l}}
\newcommand{\rS}{r_\textsc{s}}
\newcommand{\ruh}{r_\textsc{uh}}
\begin{document}
\title{Asymptotically Lifshitz spacetimes with universal horizons in $(1 + 2)$ dimensions}
\author{Sayandeb Basu}\email{sbasu@pacific.edu}
\affiliation{Physics Department, University of the Pacific, Stockton, CA 95211, USA}
\author{Jishnu Bhattacharyya}\email{jishnu.bhattacharyya@nottingham.ac.uk}
\affiliation{School of Mathematical Sciences, University of Nottingham, University Park, Nottingham, NG7 2RD, UK}
\author{David Mattingly}\email{david.mattingly@unh.edu}
\author{Matthew Roberson}\email{mkrbrson@gmail.com}
\affiliation{Department of Physics, University of New Hampshire, Durham, NH 03824, USA}
\begin{abstract}
Ho\v{r}ava gravity theory possesses global Lifshitz space as a solution and has been conjectured to provide a natural framework for Lifshitz holography. We derive the conditions on the two derivative Ho\v{r}ava gravity Lagrangian that are necessary for static, asymptotically Lifshitz spacetimes with flat transverse dimensions to contain a universal horizon, which plays a similar thermodynamic role as the Killing horizon in general relativity. Specializing to $z=2$ in $1+2$ dimensions, we then numerically construct such regular solutions over the whole spacetime. We calculate the mass for these solutions and show that, unlike the asymptotically anti-de Sitter case, the first law applied to the universal horizon is straightforwardly compatible with a thermodynamic interpretation. 
\end{abstract}

\maketitle

\section{Introduction}
Construction of holographic duals for Lifshitz field theories is an important and active line of research. Such duals release holographic approaches from the straitjacket of relativistic conformal field theory and thereby tremendously expand the types of systems holographic methods can be applied to. Any gravitational dual to a Lifshitz field theory must possess solutions that exhibit Lifshitz symmetry somewhere in the spacetime. Lifshitz geometry is not a solution of the vacuum Einstein equations, however, and so gravitational duals of Lifshitz field theories generally either possess extra tensor fields or otherwise modify the Einstein-Hilbert action of general relativity. For example, spacetimes with Lifshitz geometry somewhere in the bulk can be solutions of general relativity with extra gauge fields~\cite{Kachru:2008yh}, Einstein-Maxwell-dilaton theory~\cite{Goldstein:2009cv} and Einstein-Proca theory~\cite{Taylor:2008tg}. 

Lifshitz symmetry either asymptotically or in the bulk is not an inherent feature of any of the above theories, but merely a class of solutions. There is one gravitational theory, however, where Lifshitz symmetry is in fact intimately related to the structure of the theory: Ho\v{r}ava-Lifshitz theory, or Ho\v{r}ava gravity for short~\cite{Horava:2009uw}. Ho\v{r}ava gravity is a modified theory of gravity with a preferred foliation. The preferred foliation on the spacetime permits a splitting of spacetime into space and time in a preferred manner, thereby allowing for the imposition of a Lifshitz symmetry on the theory at high energies. This in turn renders the theory power counting renormalizable without introducing ghosts, unlike what happens in higher curvature relativistic gravity~\cite{Horava:2009uw,Visser:2009fg}\footnote{In fact,~\emph{projectable} Ho\v{r}ava gravity is perturbatively renormalizable~\cite{Barvinsky:2015kil}.}. Ho\v{r}ava gravity therefore serves as a well-behaved candidate theory of quantum gravity.

Our interest is in using Ho\v{r}ava gravity as a gravitational dual to non-gravitational Lifshitz field theories. Typically a holographic construction first requires a duality between a zero temperature field theory on the boundary and a bulk solution. Indeed, it has been argued that Ho\v{r}ava gravity on a globally Lifshitz background provides a better gravitational dual for zero temperature Lifshitz field theories, as certain quantities not reproduced in a relativistic gravitational dual naturally fall out from Ho\v{r}ava gravity when considering a global Lifshitz solution~\cite{Griffin:2012qx}.

In the usual constructions, one extends a zero temperature field theory duality to finite temperature by considering gravitational solutions containing a black hole in the bulk, with the Hawking temperature of the black hole corresponding to the temperature of the dual theory. In the Ho\v{r}ava case, however, this identification becomes immediately problematic as black hole thermodynamics in Ho\v{r}ava gravity is poorly understood. Due to the non-relativistic Lifshitz symmetry in the UV, high energy excitations in Ho\v{r}ava gravity can typically propagate faster than light. Excitations propagate towards the future relative to the preferred foliation, and hence there is a well-defined notion of causality~\cite{Bhattacharyya:2015gwa}, but UV excitations can escape from inside a Killing horizon of a static black hole solution in Ho\v{r}ava gravity. Therefore the usual Killing horizons familiar from general relativity (and extensions such as apparent horizons appropriate to more dynamic settings) no longer play the role of causal boundaries. As a consequence, there is no generic first law associated with Killing horizons~\cite{Foster:2005fr} and hence no horizon thermodynamics. Hence it is unclear how to extend any duality between the global Lifshitz solution and a zero temperature Lifshitz field theory on the boundary to finite temperature.

\begin{figure}[htb]
		\centering
		\includegraphics[scale=0.7]{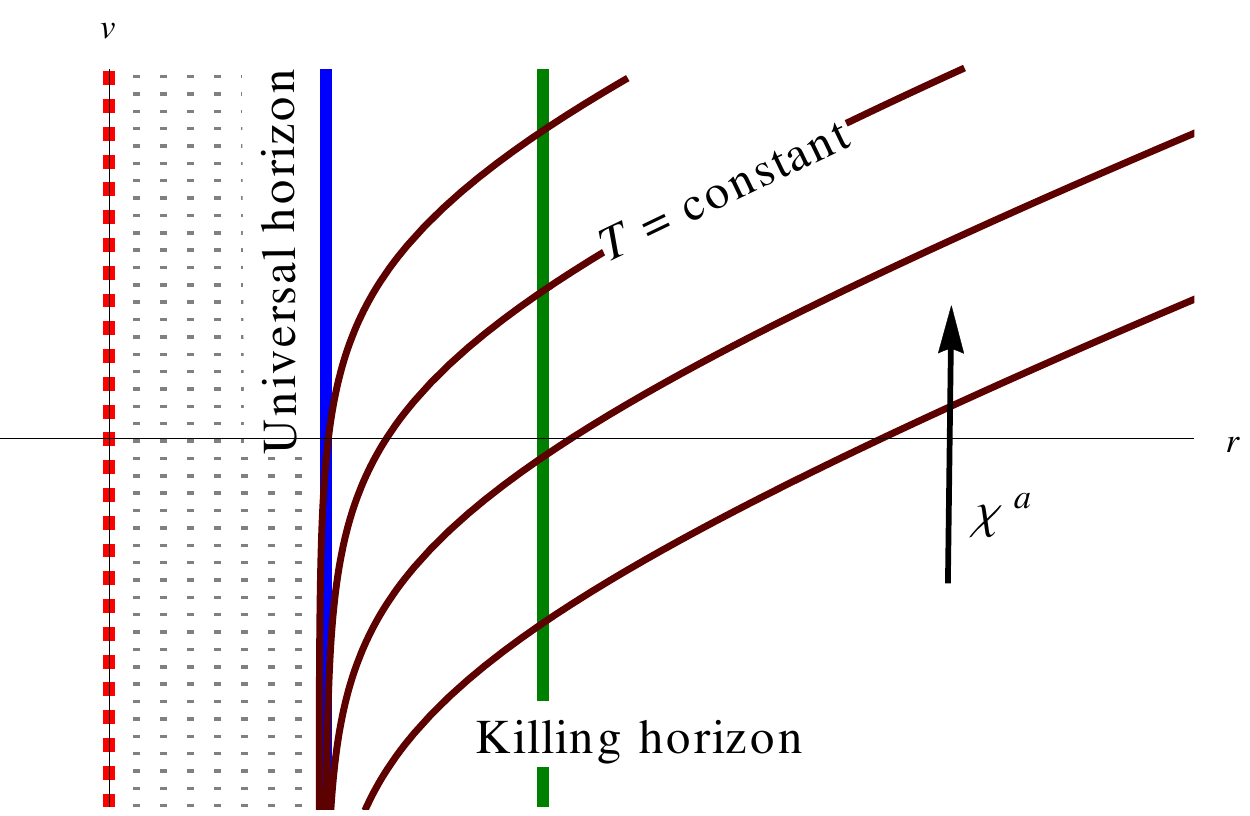}
		\caption{Bending of the preferred ($T = $ constant) hypersurfaces (thick brown lines) in ingoing Eddington-Finkelstein type coordinates in a static and spherically symmetric black hole solution of Ho\v{r}ava gravity. The Killing vector $\chi^a$ points upward throughout everywhere. The vertical green line is a constant $r$ hypersurface and denotes the usual Killing horizon defined by $\met_{a b}\chi^a\chi^b = 0$. The universal horizon of a Ho\v{r}ava gravity black hole, denoted by the vertical blue line, is also a constant $r$ hypersurface (located at $r = \ruh$) defined by the condition $u_a\chi^a = 0$, where $u_a$ is the unit timelike normal vector to the constant $T$ hypersurfaces. The dotted region inside the universal horizon (\ie for $r < \ruh$) denotes a black hole region even for arbitrary fast excitations; the constant $T$ hypersurfaces for this region are not shown to keep the diagram clean.}
	\label{fig:bendingT}
	\end{figure}

A possible prescription for establishing a finite temperature duality is provided by analyzing the physics of universal horizons. Universal horizons are the true causal boundaries of bounded bulk regions in non-projectable Ho\v{r}ava gravity~\cite{Barausse:2011pu,Blas:2011ni}. While the notion of the universal horizon can be formalized beyond any symmetries~\cite{Bhattacharyya:2015gwa}, for our present purpose it suffices to present them within the context of spherically symmetric black hole spacetimes with flat or AdS asymptotics. If we label the leaves of the preferred foliation by a scalar function $T$ and denote one such leaf by $\Sigma_T$ then each $\Sigma_T$ can bend in such a way as to still create an event horizon even for arbitrarily fast excitations, as shown in figure~\ref{fig:bendingT}. Any excitation trapped inside the universal horizon (dotted region in figure~\ref{fig:bendingT}) has to move `backward in time' with respect to the preferred foliation in order to escape to infinity, and thereby violate causality. Universal horizons have been found in $D = 1 + 2$ dimensioal Ho\v{r}ava gravity, in analogy to BTZ black holes~\cite{Sotiriou:2014gna}, in spherical symmetry with AdS, and flat asymptotics in four dimensions~\cite{Barausse:2011pu,Blas:2011ni,Bhattacharyya:2014kta}, and for the slowly rotating asymptotically flat case in four dimensions~\cite{Barausse:2012qh}.

Universal horizons do obey a first law~\cite{Berglund:2012bu,Mohd:2013zca}. Tunneling and discontinuity calculations using eternal universal horizon geometries indicate that they do radiate thermally~\cite{Berglund:2012fk,Cropp:2013sea}, although calculations in collapsing geometries give a different picture~\cite{Michel:2015rsa}. Obviously for a complete holographic construction a full thermodynamics of universal horizons must be built. In this paper we take a more modest goal: if a holographic construction for finite temperature Lifshitz field theories using Ho\v{r}ava gravity is to be constructed, we need to, at the very least, find static solutions that are asymptotically Lifshitz and contain universal (and Killing) horizons in Ho\v{r}ava gravity. These solutions, which we construct numerically, are the focus of this paper. For earlier attempts in this direction see, \eg~\cite{Janiszewski:2014iaa,Shu:2014eza}.

In order to minimize the algebraic complexity of the field equations we reduce to $1 + 2$ dimensions, although our approach is easily adaptable to higher dimensions as long as one assumes transverse planar, rather than spherical, symmetry. The reduction to $D = 1 + 2$ is not a hindrance for eventual holographic uses, as for example AdS$_3$/CFT$_2$ duality is one of the best understood implementations of holography. 

The paper is organized as follows. In section~\ref{HL}, we introduce Ho\v{r}ava gravity, the reduced action, and the relevant equations of motion. In section~\ref{Lif} we review the global Lifshitz solution in $D = 1 + 2$ dimensions and detail how the choice of Lifshitz asymptotics restricts the coefficients in the Lagrangian. We also discuss the consequence of the so called `spin-$0$ regularity' in this section. In section~\ref{LifBH} we describe our numeric procedure and give some example solutions. The corresponding Smarr formulae and first laws are presented in section~\ref{Smarr}. Finally, we summarize in the conclusions~\ref{Conclusions}. Throughout the paper we use metric signature $(-,+,+)$.
\section{Ho\v{r}ava gravity and equations of motion}~\label{HL}
\subsection{The action and equations of motion}
Ho\v{r}ava gravity can be covariantly formulated as a scalar-tensor theory, where the dynamical scalar field $T$, called the~\emph{khronon}, always admits a non-zero timelike gradient everywhere on-shell. This allows one to construct a unit-timelike hypersurface orthogonal one-form $u_a$, called the~\emph{{\ae}ther}, such that
	\beql{ae:HSO:norm}
	u_a = -N\Dl_a T, \qquad \met^{a b}u_a u_b = -1~,
	\eeq
where the function $N$ is solved for via the unit norm constraint as follows 
	\beql{eq:lapse}
	N^{-2} = -\met^{a b}(\Dl_a T)(\Dl_b T)~.
	\eeq
Besides the usual diffeomorphism, Ho\v{r}ava gravity is also invariant under arbitrary reparametrizations of the khronon: $T \mapsto \tilde{T} = \tilde{T}(T)$. Under such reparametrizations  $N$ is required to transform as $N \mapsto \tilde{N} = (\ed\tilde{T}/\dT)^{-1}N$, such that the {\ae}ther remains manifestly invariant under the reparametrizations of the khronon. This allows one to express the (two-derivative truncated/IR limit) action of Ho\v{r}ava gravity in $D = (1 + 2)$ dimensions in a manifestly covariant and reparametrization invariant manner as follows~\cite{Sotiriou:2011dr}\footnote{The complete action of Ho\v{r}ava gravity can also be covariantized via such `St\"uckelbering' procedure~\cite{Sotiriou:2011dr}. In this work, however, we only work with the IR limit of the theory.}
	\beql{action}
	S = \frac{1}{16\pi G_{\ae}}\int\ed^3x\sqrt{-\met}(-2\Lambda_{\CC} + R + \lag) + S_{\GHY} + S_{T, b}.
	\eeq
Here $\Lambda_{\CC}$ is the cosmological constant which will be taken to be negative in this work, $R$ is the curvature scalar, and $\lag$ is the khronon's Lagrangian  given by
	\beql{lag:ae}
	\lag = -\tensor{Z}{^{a b}_{c d}}(\Dl_a u^c)(\Dl_b u^d)~.
	\eeq
The tensor $\tensor{Z}{^{a c}_{c d}}$ is given by
	\beql{def:Zabcd}
	\tensor{Z}{^{a b}_{c d}} = c_1\met^{a b}\met_{c d} + c_2\tensor{\delta}{^a_c}\tensor{\delta}{^b_d} + c_3\tensor{\delta}{^a_d}\tensor{\delta}{^b_c} - c_4u^au^b\met_{c d}~,
	\eeq
where $c_1$, $c_2$, $c_3$, $c_4$ are coupling constants. $S_{\GHY}$ is the standard Gibbons-Hawking-York boundary term and $S_{T,b}$ represents any additional boundary terms necessary due to the presence of the khronon field. We will return to the boundary terms in section~\ref{sec:nustar} when we discuss the total mass of solutions, but these boundary terms are irrelevant for a derivation of the bulk equations of motion. The bulk covariant equations of motion for the metric and khronon are generated by extremizing the action~\eqref{action} under variations of the respective fields, with the assumption that the {\ae}ther is derived from the khronon via~\eqref{ae:HSO:norm}.

The khronon equation of motion and corresponding solutions of Ho\v{r}ava gravity are most efficiently analyzed by leveraging the relation between Ho\v{r}ava gravity and Einstein-{\ae}ther theory, as we now explain. The action~\eqref{action}, with the {\ae}ther~\emph{only} satisfying the unit norm constraint (\ie not hypersurface orthogonality) and being itself treated as the fundamental field, leads to Einstein-{\ae}ther theory~\cite{Jacobson:2000xp}, a vector-tensor theory of gravity coupled to a unit timelike vector field. One may subsequently restrict attention to the hypersurface orthogonal sector of Einstein-{\ae}ther theory by imposing the hypersurface orthogonality condition on the {\ae}ther~\eqref{ae:HSO:norm} at the level of the equations of motion. Neglecting all boundary terms, the Einstein's equations generated by extremizing the action~\eqref{action} under variations of the metric, leads to formally identical Einstein's equations for both Ho\v{r}ava gravity and the hypersurface orthogonal sector of Einstein-{\ae}ther theory~\cite{Jacobson:2010mx} (see also~\cite{Bhattacharyya:2015uxt} for a more recent discussion, especially from the perspective of the initial value problem in both theories). However, the corresponding bulk {\ae}ther equations of motion in Einstein-{\ae}ther theory is
	\beql{EOM:ae}
	\vAE^a = 0~,
	\eeq
where $\vAE^a$ is the `component' of the functional derivative of the action~\eqref{action} with respect to the {\ae}ther which is orthogonal to the {\ae}ther itself (\ie $u_a\vAE^a = 0$), while the khronon's equations of motion in Ho\v{r}ava gravity reads 
	\beql{EOM:HL}
	\Dl_a[N\vAE^a] = 0~.
	\eeq
The formal equivalence of the Einstein's equations, taken together with the similarities of~\eqref{EOM:ae} and~\eqref{EOM:HL}, make it clear that any solution of the hypersurface orthogonal sector Einstein-{\ae}ther theory is also a solution of Ho\v{r}ava gravity~\cite{Jacobson:2010mx}, although the converse is generally not true.

In this work, we will restrict ourselves to static solutions of Ho\v{r}ava theory in $D = (1 + 2)$ with translational symmetry in the transverse space (see below). In a similar setting, the {\ae}ther in Einstein-{\ae}ther theory is automatically hypersurface orthogonal as dictated by the symmetries. One may then argue along the lines of~\cite{Bhattacharyya:2014kta} to conclude that solutions of Ho\v{r}ava theory with these symmetries, and~\emph{admitting a regular universal horizon} in addition, are also the only solutions of Einstein-{\ae}ther theory with these properties (note that the asymptotic behaviour of the solutions is irrelevant in this argument). Therefore, it suffices to solve the Einstein-{\ae}ther equations of motion to obtain the desired solutions in Ho\v{r}ava gravity; this will be the approach taken in this paper.

Even though the individual couplings $c_1, \cdots, c_4$ appear directly in the action~\eqref{action}, one may argue that owing to the hypersurface orthogonality of the {\ae}ther, only the combinations $c_{13} = c_1 + c_3$, $c_{123} = c_2 + c_{13}$ and $c_{14} = c_1 + c_4$ show up explicitly in all subsequent expressions~\cite{Barausse:2011pu}. Finally, it will be useful to note the following kinematical quantities: $a^a = u^b\Dl_b u^a$ being the acceleration of the {\ae}ther congruence, $K_{a b} = \Dl_a u_b + u_a a_b$ being the extrinsic curvature of the constant khronon hypersurfaces, and $K = \met^{a b}K_{a b}$ being the corresponding mean curvature.
\subsection{Equations of motion under staticity and transverse space translation symmetry}
It will be convenient to use ingoing Eddington-Finkelstein type (EF) coordinates, in which the metric on a static spacetime, with translational symmetry in the transverse space, becomes
	\beql{met:EF}
	\ds^2 = -e(r)\dv^2 + 2f(r)\dv\dr + r^2\dy^2,
	\eeq
where $r$ is the canonical radial coordinate, and $y$ is the coordinate on the transverse space. Note that $y$ is not a bounded coordinate; rather $-\infty < y < \infty$. The Killing vector associated with staticity, denoted by $\chi^a$, is given by $\chi^a = \dl_v$ in these coordinates, while the Killing vector associated with the translational symmetry in the transverse space (\ie under $y \to y + $ constant) is $\dl_{y}$.

The {\ae}ther one-form decomposes in these coordinates as
	\beql{ae:EF}
	u_a = (u\cdot\chi)\dv + \frac{f(r)\dr}{(s\cdot\chi) - (u\cdot\chi)}~,
	\eeq
where $(s\cdot\chi) = s_a\chi^a$, $s^a$ being the unique (`outward pointing') spacelike unit vector which is orthogonal to both the {\ae}ther and the transverse direction. As already mentioned, the symmetries of the spacetime make the {\ae}ther hypersurface orthogonal as the above expression also manifestly reveals (the functions $(u\cdot\chi)$ and $(s\cdot\chi)$ are functions of $r$ only), while the unit-norm constraint on the {\ae}ther~\eqref{ae:HSO:norm} is taken into account via
	\beql{def:e}
	e(r) = (u\cdot\chi)^2 - (s\cdot\chi)^2~.
	\eeq
The functions $e(r)$ and $f(r)$ capture the free metric components that one needs to solve for from the equations of motion of Ho\v{r}ava gravity. The (symmetry reduced) {\ae}ther has one additional free component. It will be algebraically beneficial to write this component via the variable $X(r)$ defined by
	\beql{def:X}
	X(r) = (s\cdot\chi) - (u\cdot\chi)~.
	\eeq
In what follows, the equations of motion will be solved for the functions $e(r)$, $X(r)$ and $f(r)$ for reasons to be explained below, and the functions $(u\cdot\chi)$ and $(s\cdot\chi)$ can then be determined by inverting~\eqref{def:e} and~\eqref{def:X}.

Instead of adapting the fully general covariant equations of motion to the above symmetries, it is more convenient to substitute the above symmetry-adapted expressions for the metric and the {\ae}ther into the action~\eqref{action} directly, which yields the following time independent action
	\begin{multline}\label{eq:tiaction}
	S = -\int\dr (4 r f^2 X^4)^{-1} \\
	[-2 e X (f (r^2 (c_{123} - c_{14}) e' X'  \\
	-r X (r (c_{123} + c_{14}) X'^2+c_2 e')+ c_{123} X^3)\\
	+4 r X^3 f')+X^2 (f (-2 r^2 X (c_{123} + c_{14}) e' X'\\
	+r^2 (c_{123} - c_{14}) e'^2+r X^2 (r (c_{123} - c_{14}) X'^2\\
	-2 (c_2-4) e'+4 r e'')+ c_{123}X^4+2 c_2 r X^3 X')\\
	-4 r^2 X^2 e' f'+8 \Lambda_{\CC}  r^2 f^3 X^2)+e^2 f (r^2 (c_{123} - c_{14}) X'^2\\
	+ c_{123}X^2-2 c_2 r X X')]~,
	\end{multline}
where $'$ denotes differentiation with respect to $r$. The equations of motion are then generated by extremizing the above time independent action with respect to variations of the three independent free components of the metric and the {\ae}ther: $e(r)$, $f(r)$, and $X(r)$. While solutions to the equations thus obtained are not always guaranteed to be static solutions of the original covariant equations of motion, the set of solutions of the equations from~\eqref{eq:tiaction}~\emph{is} guaranteed to include solutions of the full covariant equations. In other words, being a solution of the equations from~\eqref{eq:tiaction} is a necessary but not sufficient condition on static solutions of the covariant equations. We will therefore look for solutions of the equations from~\eqref{eq:tiaction} and then check if they are static solutions of the covariant equations of motion~\emph{a posteriori}.

The equations for $e(r)$, $f(r)$, and $X(r)$ following from~\eqref{eq:tiaction} are rather complicated coupled ODEs and are not particularly illuminating, so we will not reproduce them if full here. However, there are some important structural aspects that need to be mentioned. First, one may note note that the time independent action~\eqref{eq:tiaction} does not contain any term that is quadratic in derivatives of $f(r)$, either via $f''(r)$ or $f'(r)^2$. As a result, the equation of motion for $f(r)$ is an algebraic equation, which simplifies the system considerably (and is the primary motivation for choosing $X(r)$ as a fundamental free component). In our subsequent numerical analysis, we substitute this algebraic expression for $f(r)$ back into the equations of motion for $e(r)$ and $X(r)$ which yields two second order differential equations for $e''(r)$ and $X''(r)$ in terms of $e(r)$, $e'(r)$, $X(r)$ and $X'(r)$.

Second, the resulting differential equations for $e(r)$ and $X(r)$ both na\"ively have a singularity at a particular value of the pair $e(r)$ and $X(r)$. The source of this singularity is a feature previously found in studies of black holes in Ho\v{r}ava gravity known as the~\emph{spin-$0$ horizon}. In the present setting, unlike general relativity in $D = (1 + 2)$, Ho\v{r}ava gravity is known to contain a propagating scalar or spin-$0$ mode with local (low energy) speed $s_0$ relative to the {\ae}ther frame given by the expression~\cite{Sotiriou:2011dr} (compare with the corresponding expression in $D = 1 + 3$~\cite{Jacobson:2004ts})
	\beql{speed:s0}
	s_0^2 = \frac{c_{123}}{c_{14}(1 - c_{13})(1 + c_{13} + 2c_2)}~.
	\eeq
The different local speed relative to the {\ae}ther frame is equivalently described by stating that the low energy spin-$0$ mode propagates on the light cone of an effective~\emph{spin-$0$ metric} $\met^{(0)}_{a b}$ given by
	\beql{met:s0}
	\met^{(0)}_{a b} = \met_{a b} - (s_0^2 - 1)u_a u_b.
	\eeq
The low energy spin-$0$ mode has a corresponding causal horizon, known as the~\emph{spin-$0$ horizon}, and its radial location is given by the largest root of $|\chi|_{s_0}^2 \equiv \met^{(0)}_{a b}\chi^a\chi^b = 0$, analogous to the Killing horizon in general relativity. On this horizon the equations of motion break down (\emph{c.f.} the discussion in~\cite{Eling:2006ec}). In our case, this is reflected in the equations of motion for $e(r)$ and $X(r)$ which take the form
	\begin{eqnarray}\label{eq:poles}
	e''(r) = \frac{F_e(e, e', X, X', r, c_i)}{8 (1 - c_{13} c_{14} (1 + c_{13} + 2c_2)) r^2 X(r)^6|\chi|_{s_0}^2} \\
	X''(r) = \frac{F_X(e, e', X, X', r, c_i)}{8 (1 - c_{13} c_{14} (1 + c_{13} + 2c_2)) r^2 X(r)^5|\chi|_{s_0}^2}.
	\end{eqnarray}
$F_e$ and $F_X$ are complicated and unilluminating functions and hence their full form will be omitted. On the spin-$0$ horizon this equation will generally be unstable unless $F_e$ and $F_X$ also vanish. This regularity requirement will eventually reduce our black hole solutions down to a one parameter family. We will return to this issue when we describe our numerical approach.

The spin-$0$ horizon has a useful property in that it can be `moved around' relative to a Killing horizon via a field redefinition. As noted in~\cite{Foster:2005ec}, under disformal field redefinitions, \ie redefinitions of the form 
	\beql{eq:disformal}
	\met'_{a b} = \met_{a b} - (1 - \sigma^2) u_a u_b~, \quad u^a = \sigma^{-1}u^a, \quad \sigma > 0,
	\eeq
the action~\eqref{action} transforms into itself with simply new values of the $c_i$ coefficients. In particular, the coefficients transform as
	\beql{eq:coeffdisformal}
	1 - c'_{13} = \sigma(1 - c_{13}), \quad c'_{123} = \sigma c_{123}, \quad c'_{14} = c_{14}.
	\eeq
The speed of the spin-$0$ mode is not invariant under the disformal redefinitions, and in fact, given an initial set of coefficients one can always perform a field redefinition such that the spin-$0$ speed becomes unity. In other words, one can always set the Killing horizon and spin-$0$ horizon to be co-located without loss of generality. This will simplify the numerical analysis.

In the present work, we wish to seek solutions of Ho\v{r}ava gravity with Lifshitz asymptotics and a regular universal horizon in the bulk. In particular, we need to solve~\eqref{eq:poles} with asymptotically Lifshitz boundary conditions on the metric and {\ae}ther components. To that end, we need to derive the appropriate asymptotic behaviour of the functions $e(r)$, $X(r)$ and $f(r)$ as $r \to \infty$, as well as the conditions under which the solutions are also regular in the bulk of the spacetime, especially on their respective spin-$0$ horizons. These issues will be taken up in the following section, which will also pave the way towards the numerical construction of the sought after solutions.
\section{Asymptotically Lifshitz spacetimes}\label{Lif}
\subsection{The global Lifshitz solution}
Before we can properly discuss asymptotically Lifshitz spacetimes we first must discuss the global background Lifshitz solution which plays the same role global AdS space does for asymptotically AdS spacetimes. In  $D = 1 + 2$ dimensions in the canonical (Schwarzschild-type) $t$, $r$ and $y$ coordinates ($y$ being the transverse coordinate), the global Lifshitz spacetime introduced in~\cite{Kachru:2008yh} is an obvious generalization of AdS$_3$ spacetime, but with inhomogeneous scale invariance between space and time. In its standard/canonical form, the metric of the global Lifshitz spacetime in $D = 1 + 2$ is
	\beql{met:global}
	\ds^2 = -(r/\elL)^{2z}\dt^2 + (r/\elL)^{-2}\dr^2 + (r/\elL)^2\dy^2~,
	\eeq
where the constant $z \geqq 1$ is the (Lifshitz) scaling exponent and the fixed length scale $\elL$ is the Lifshitz scale. For $z = 1$, the metric~\eqref{met:global} describes AdS$_3$.

The metric~\eqref{met:global} is manifestly isometric under constant translations of the time coordinate $t$, under $t \mapsto -t$, as well as under constant translations of the transverse space coordinate $y$. More interestingly, the metric~\eqref{met:global} is also isometric under scale transformations of the form 
	\beql{scaling}
	t \mapsto \lambda^z\,t, \qquad y \mapsto \lambda\,y, \qquad r \mapsto \lambda^{-1}r~.
	\eeq
Clearly for $z > 1$,  the scale invariance between $t$ and $y$ is~\emph{inhomogeneous}.

We are eventually interested in constructing black hole solutions which are only asymptically Lifshitz, and for that purpose it will be useful to switch to ingoing EF coordinates~\eqref{met:EF}. In particular, the metric~\eqref{met:global} of the global Lifshitz spacetime takes the following form in EF coordinates (compare with~\eqref{met:EF}),
	\beql{met:global:EF}
	\ds^2 = -(r/\elL)^{2z}\dv^2 + 2(r/\elL)^{z - 1}\dv\dr + (r/\elL)^2\dy^2~.
	\eeq
The scale transformations analogous to~\eqref{scaling} leaving the metric~\eqref{met:global:EF} invariant are\footnote{This follows from the definition of the $v$ coordinate: $\dv = \dt + (r/\elL)^{-(z + 1)}\dr$.}
	\beql{scaling:gen:EF}
	v \mapsto \lambda^z v~, \qquad y \mapsto \lambda y~, \qquad r \mapsto \lambda^{-1}r~.
	\eeq
As discovered in~\cite{Griffin:2012qx}, the global Lifshitz metric~\eqref{met:global:EF} is a solution to the Ho\v{r}ava gravity equations of motion, along with the following profile for the {\ae}ther (compare with~\eqref{ae:EF})
	\beql{u_a:global:EF}
	u_a = -(r/\elL)^{z}\dv + (r/\elL)^{-1}\dr~,
	\eeq
In particular, the {\ae}ther satisfies (as per requirement) all the above symmetries including that under~\eqref{scaling:gen:EF} and is aligned with the Killing vector $\chi^a$ everywhere\footnote{It can be easily proved that in a globally Lifshitz solution, the equations of motion of Ho\v{r}ava gravity forces the {\ae}ther to be globally aligned with the Killing vector $\chi^a$.}. The solution parameters $z$ and $\elL$ are fully determined by the parameters $\Lambda_{\CC}$ and $c_{14}$ by the following relations
	\beql{global:CC-c14}
	\Lambda_{\CC} = -\frac{z(z + 1)}{2\elL^2}~, \qquad c_{14} = \frac{z - 1}{z}~.
	\eeq
The second relation means, in particular, that the Lifshitz exponent is uniquely determined by the coupling $c_{14}$. Notice that the global solution is independent of the values of the couplings $c_{13}$ and $c_2$.
\subsection{Asymptotic expansion}\label{sec:asymptotics}
Moving on to static, asymptotically Lifshitz spacetimes, it is not immediately clear under what conditions the various metric and {\ae}ther coefficients admit a well-defined power series in $r^{-1}$; this is a concern especially when $z$ is non-integer. We must therefore construct a useful parametrization of the asymptotic forms of the various metric and {\ae}ther coefficients around $r = \infty$. 

While in the global Lifshitz solution the {\ae}ther is globally aligned with the Killing vector $\chi^a$, in an asymptotically Lifshitz case this will not be the case everywhere in the spacetime. Rather, we merely require an asymptotic alignment between the {\ae}ther and $\chi^a$. The additional measure for the `misalignment' between the {\ae}ther and $\chi^a$ is conveniently captured through the quantity $(s\cdot\chi) \equiv s_a\chi^a$, where $s^a$, as introduced previously is the unique outwards pointing unit spacelike vector orthogonal to the {\ae}ther and the transverse directions everywhere. Intuitively, we wish to define an asymptotically Lifshitz spacetime in the present context as a spacetime where the {\ae}ther becomes aligned with the Killing vector and the metric approaches the global Lifshitz solution as $r \to \infty$. These conditions can be properly implemented in the present coordinates by requiring
	\beql{def:asymp-Lifshitz}
	\begin{split}
	\lim_{r \to \infty}\frac{(u\cdot\chi)}{\sqrt{-\chi\cdot\chi}} & = -1~, \\
	\lim_{r \to \infty}\frac{(s\cdot\chi)}{\sqrt{-\chi\cdot\chi}} & = 0~, \\
	\lim_{r \to \infty}\frac{f(r)}{r^{z - 1}} & = 1.
	\end{split}
	\eeq
Since we wish our solutions to smoothly approach the global solution upon tuning some parameters (\eg the mass) we can factor out the appropriate global Lifshitz behaviours from $(u \cdot \chi)$, $e(r) \equiv -(\chi\cdot\chi)$, and $f(r)$ and write
	\beql{def:E0-U0}
	\begin{split}
	        e(r) & = (r/\elL)^{2z}E_0(r)~,  \\
	(u\cdot\chi) & = -(r/\elL)^{z}U_0(r)~,\\
	f(r) & = (r/\elL)^{z - 1}F_0(r)~,
	\end{split}
	\eeq
such that the conditions~\eqref{def:asymp-Lifshitz} becomes equivalent to
	\beql{def:asymp-Lifshitz:EUF}
	\lim_{r \to \infty} E_0(r) = 1~, \quad \lim_{r \to \infty} U_0(r) = 1~, \quad \lim_{r \to \infty} F_0(r)= 1~.
	\eeq
Asymptotically Lifshitz spacetimes are not, of course,~\emph{necessarily} solutions of the Ho\v{r}ava gravity equations of motion. Rather, for asymptotically Lifshitz solutions, the functions $U_0(r)$, $E_0(r)$ and $F_0(r)$ not only must have well-defined limits to $r = \infty$ but also must satisfy an asymptotic expansion of the equations of motion. As we shall see, the asymptotic equations of motions yield a significant restriction on the choice of the $c_i$ coefficients.

In order to compute the asymptotic equations of motion we need some convenient parametrization of the fall-offs of these functions as $r\rightarrow \infty$. To that end, we will~\emph{assume} that~\emph{given some $z$, there exists a number $\nu_{\star} > 0$ such that the functions $U_0(r)$, $E_0(r)$, and $F_0(r)$ are analytic at $r = \infty$ in $r^{-\nu_{\star}}$, \ie, they all admit well defined power series (albeit asymptotic) expansions in powers of $r^{-\nu_{\star}}$ as follows:}
	\beql{series:E0-U0-F0}
	\begin{split}
	E_0(r) & = 1 + \frac{e_1(z + 1)\elL^{\nu_{\star}}}{r^{\nu_{\star}}} + \frac{e_2(z + 1)^2\elL^{2\nu_{\star}}}{r^{2\nu_{\star}}} + \ord(r^{-3\nu_{\star}})~, \\
	U_0(r) & = 1 + \frac{u_1(z + 1)\elL^{\nu_{\star}}}{r^{\nu_{\star}}} + \frac{u_2(z + 1)^2\elL^{2\nu_{\star}}}{r^{2\nu_{\star}}} + \ord(r^{-3\nu_{\star}})~, \\
	F_0(r) & = 1 + \frac{f_1(z + 1)\elL^{\nu_{\star}}}{r^{\nu_{\star}}} + \frac{f_2(z + 1)^2\elL^{2\nu_{\star}}}{r^{2\nu_{\star}}} + \ord(r^{-3\nu_{\star}})~.
	\end{split}
	\eeq
In particular, the $\ord(r^{-n\nu_{\star}})$ coefficient for some integer $n \geqq 1$ has been defined with an explicit factor of $(z + 1)^n$ for convenience with the asymptotic analysis, as well as a factor of $\elL^{n\nu_{\star}}$ has been included to make the coefficients dimensionless. 

The above expansions yield analogous expansion for $(s\cdot\chi)$ and $X(r)$ via~\eqref{def:e} and~\eqref{def:X}. For the expansion of $(s\cdot\chi)$ in particular, we may start with the following expression
	\beqn{
	(s\cdot\chi)^2 = (r/\elL)^{2z}\left[U_0(r)^2 - E_0(r)\right]~,
	}
which follows from~\eqref{def:e}. If we plug in the ansatz~\eqref{series:E0-U0-F0} above, we end up with the following asymptotic behaviour for $(s\cdot\chi)$
	\beqn{
	(s\cdot\chi)^2 = (r/\elL)^{2z}\ord(r^{-n_s\nu_{\star}})~,
	}
which encompasses the possibility that, for some integer $n_s \geqq 1$, the first $(n_s - 1)$ terms in the series for $U_0(r)^2 - E_0(r)$ are zero. In other words, $(s\cdot\chi)$ may have the following asymptotic behaviour
	\beql{def:S0}
	(s\cdot\chi) = (r/\elL)^{z - \frac{n_s\nu_{\star}}{2}}S_0(r)~, 
	\eeq
along with
	\beql{series:S0}
	S_0(r) = c_{\ae} + \frac{s_1(z + 1)\elL^{\nu_{\star}}}{r^{\nu_{\star}}} + \frac{s_2(z + 1)^2\elL^{2\nu_{\star}}}{r^{2\nu_{\star}}} + \ord(r^{-3\nu_{\star}})~,
	\eeq
such that for a given $n_s$, $c_{\ae} \neq 0$ is a constant which captures the leading order behaviour of $S_0(r)$. In particular, the case $n_s = 0$ is not allowed on account of the presumed asymptotically Lifshitz behaviour; indeed, for all $n_s \geqq 1$ one finds that the second condition in the definition of asymptotic Lifshitz-ness proposed in~\eqref{def:asymp-Lifshitz} is also met. This expansion for $(s\cdot \chi)$  also generates an asymptotic expansion for $X(r)$ via~\eqref{def:X}.
\subsection{Boundaries, mass and determination of $\nu_{\star}$}\label{sec:nustar}
To proceed in the expansion we need to determine $\nu_{\star}$, which can be done by requiring a non-zero but finite mass for the black hole solutions. Unlike the local equations of motion, the total mass~\emph{does} depend on the boundary terms present in the action~\eqref{action}. Therefore, the first step is to deal with the additional possible boundary term $S_{T,b}$.

Boundary terms are introduced into actions so that the variational principle is well defined. The variation of the GHY term, for example, explicitly cancels the boundary term generated when varying the Einstein-Hilbert term and imposing Dirichlet boundary conditions on the metric. Since Ho\v{r}ava gravity has the Einstein-Hilbert term in the bulk action~\eqref{action}, the GHY term is necessary if we maintain Dirichlet boundary conditions for the metric. We must check, however, if a) the other terms in the Ho\v{r}ava gravity action are compatible with Dirichlet metric conditions, b) what type of boundary conditions are appropriate for the khronon, and c) if additional boundary terms are generated from the khronon variation.

The variation of the bulk Ho\v{r}ava gravity action~\eqref{action} yields the following additional boundary variations 
	\beql{eq:HLboundary}
	\delta S_b = \int\limits_{\dl V} d^2x\sqrt{h}\left[n^c B_{abc} \delta \met^{a b} + n^c B_c \delta T + n^c \tensor{B}{_c^a}(\vDl_a \delta T)\right],
	\eeq
where $n^a$ is the normal to the boundary $\dl V$, $h_{ab}$ is the induced metric on the boundary, $\vDl_a$ is the projected spatial covariant derivative on the preferred foliation, and $B_{a b c}$, $\tensor{B}{_a^b}$ and $B_a$ are tensors built out of $\met_{a b}$, $h_{a b}$, $u_a$ and their derivatives. We immediately see that with Dirichlet boundary conditions for the metric the first term vanishes and hence the boundary analysis for the metric proceeds exactly as it does in general relativity: addition of the GHY term and Dirichlet boundary conditions for the metric makes the variational principle well defined for metric variations. Therefore the particular (complicated) expression for $B_{a b c}$ is irrelevant for our subsequent discussions and we will omit it.

The khronon variation is more subtle, as we have boundary variations in~\eqref{eq:HLboundary} that involve both direct variations of the khronon and also derivatives of the variations. Insight can be gained by examining what constitutes the boundary $\dl V$, as well the structure of $B_a$ and $\tensor{B}{_a^b}$, which are given by
	\beql{def:B}
	B_a = -2N \vec{\AE}_a~, \qquad \tensor{B}{_c^a} = 2N\tensor{Z}{^{a b}_{c d}}\Dl_b u^d~,
	\eeq
where $\tensor{Z}{^{a b}_{c d}}$ is defined in~\eqref{def:Zabcd}. In the simplest setting with a spacetime without any horizons and/or singularities, $\dl V$ consists of the boundary at (spatial) infinity to be denoted by $\mathscr{I}$ in what follows\footnote{Due to the modified causal struture of spacetimes in Ho\v{r}ava gravity, the boundary at spatial infinity is the only relevant boundary at infinity; see~\cite{Bhattacharyya:2015gwa} for further details.}, as well as the boundaries at infinite past and future. Since we need to adapt to the preferred foliation, the boundaries at infinite past and future are also slices of the preferred foliation. Therefore, given the form of $B_a$~\eqref{def:B}, the contribution of the second term in~\eqref{eq:HLboundary} vanishes on the boundaries at infinite past and future, since $n^a = u^a$ on these surfaces. On the other hand, the field configuration~\emph{on} $\mathscr{I}$ is that of the corresponding global solution, here Lifshitz, which satisfies the Einstein-{\ae}ther theory equations of motion. Hence the asymptotic (Lifshitz) boundary condition implies $\vec{\AE^a} = 0$ rather than $\Dl_a(N\vec{\AE}^a) = 0$. Consequently, there is never a boundary contribution from the second term in~\eqref{eq:HLboundary} for the present choice of $\dl V$.

The lack of a boundary term proportional to $\delta T$ matches the intuition gained by considering the fundamental reparametrization invariance of Ho\v{r}ava gravity. On the boundary, such variations can always be absorbed by leveraging the reparametrization invariance. Consequently, setting Dirichlet boundary conditions on the khronon is inappropriate. Rather, Neumann boundary condition is the appropriate type of boundary condition for the khronon. We are requiring our spacetimes to be asymptotically Lifshitz, which in turn puts a condition on $u^a$ at infinity; namely, it aligns with the asymptotic Killing vector that generates stationarity on $\mathscr{I}$. Since $u_a$ is related to the gradient of $T$~\eqref{ae:HSO:norm}, such Dirichlet conditions on $u^a$ correspond to Neumann conditions on the khronon. The appropriate condition to maintain $u^a$ orientation at infinity (\ie on $\mathscr{I}$) is, in fact, precisely that the spatial gradient of the khronon variation vanishes, as non-zero spatial gradients are exactly what would `tilt' $u^a$. Therefore, we impose Neumann conditions on the khronon, in particular require $\vec{\Dl_a} \delta T = 0$ on $\mathscr{I}$, which kills the third boundary term in~\eqref{eq:HLboundary}. In summary, at least for the kind of boundaries we have considered so far, the only boundary term necessary in our construction is the usual GHY term since the appropriate physical boundary conditions are Dirichlet for the metric variations and Neumann for the khronon variations.

Things are more sutle if the spacetime admits a universal horizon. In this case, imposing Dirichlet boundary condition on the metric and Neumann boundary condition on the khronon on every boundary surface still suffices to kill the first and the last terms in~\eqref{eq:HLboundary}, thereby saving us from introducing additional boundary terms in the action. However, the universal horizon raises the possibility of additional `inner boundaries' in the spacetime on which the contribution from the middle term in~\eqref{eq:HLboundary} is not necessarily zero at first sight. To resolve this, let us take a closer look at universal horizons in asymptotically Lifshitz spacetimes.

In a stationary spacetime with~\emph{flat} asymptotics, the universal horizon is a leaf of the preferred foliation that barely fails to reach the boundary at infinity~\cite{Bhattacharyya:2015gwa}. Any preferred slice that reaches spatial infinity never crosses the universal horizon, but instead asymptotes to it. The causal structure is intuitively much more accessible in spherically symmetric spacetimes, where the high degree of symmetry allows one to appeal to \eg figure~\ref{fig:bendingT} and conclude that the universal horizon has to be a leaf of the preferred foliation which is simultaneously a constant $r$ hypersurface and therefore orthogonal to the Killing vector of stationarity $\chi^a$. In other words, the universal horizon is locally characterized by the condition
	\beql{def:UH}
	(u\cdot\chi)_\textsc{uh} = 0~,
	\eeq
whose radial location will be denoted by $\ruh$\footnote{In a spacetime with multiple (disconnected) surfaces satisfying $(u\cdot\chi) = 0$, the outermost one is the universal horizon.}. The above argument can be made more rigorous, and the condition~\eqref{def:UH} still suffices as a local characterization of the universal horizon in the most general stationary spacetimes, as long as the quantity $(a\cdot\chi) \neq 0$ on the said surface~\cite{Bhattacharyya:2015gwa}.

In spacetimes with Lifshitz asymptotics, since the Killing vector $\chi^a$ is timelike asymptotically, we have the desired asymptotic behaviour of the spacetime to utilize the settings of~\cite{Bhattacharyya:2015gwa}. Additionally, the quantity $(a\cdot\chi) \neq 0$ on the universal horizon (see figure~\ref{fig:adotchivsruh}). Hence, condition~\eqref{def:UH} also provides the suitable local characterization of the universal horizon here, and the causal structure of the spacetimes we are dealing with is still qualitatively as captured in figure~\ref{fig:bendingT}.

In an asymptotically Lifshitz spacetime with a universal horizon (just as in the corresponding case of an asymptotically flat spacetime), one may divide up the spacetime into two (causally) disjoint regions, namely the `outside region' which is the part of the spacetime that is continuously connected to the boundary at infinity $\mathscr{I}$~\emph{and} where $(u\cdot\chi) < 0$ holds everywhere, and the `inside region' which is the complement of the `outside region'. The boundary of the `outside region' then consists of the boundary at infinity $\mathscr{I}$, the boundaries at infinite past and future (for the `outside region'), and the universal horizon denoting an inner boundary for the `outside region'. One may invoke our previous logic to conclude that~\eqref{eq:HLboundary} vanish on the boundary at infinity, as well as the boundaries at infinite past and future, if Dirichlet and Neumann boundary conditions are imposed on the metric and the khronon, respectively. More importantly, the (future) universal horizon coincides with the boundary at infinite future, as can be inferred \eg from figure~\ref{fig:bendingT}, and hence~\eqref{eq:HLboundary} vanishes here as well. For the `inner region', at least in the present setting, one may at most have a sequence of `inner horizons' which are themselves leaves of the preferred foliation characterized by the condition $(u\cdot\chi) = 0$ (neither of which are universal horizons however). Since these surfaces are leaves of the preferred foliation themselves, \ie $n^a = u^a$ on each of them, the boundary variation~\eqref{eq:HLboundary} vanishes on every possible inner boundaries as well. Therefore, even for the case of interest, the only boundary term necessary in our construction is the usual GHY term with Dirichlet boundary condition on the metric variations and Neumann boundary condition for the khronon variations.

Now that the question of boundary terms has been settled, we can proceed with calculating the mass. Since the GHY term is the only term, the total mass $M$ of a Lifshitz black hole solution using the preferred foliation is given by the familiar Hawking-Horowitz formula~\cite{Hawking:1995fd},
	\beql{eq:HH}
	M = -\frac{1}{8\pi G_{\ae}}\int\limits_{\mathscr{B}}\left[N\hat{k} - N^a p_{ab} n^b\right] - M_{\textsc{gl}}~,
	\eeq
where $\mathscr{B}$ is the suitable `one-boundary' -- cross-sections of the boundary $\dl V$ -- on which the `surface terms' in the Hamiltonian (here generated purely from the GHY term) contribute, $N^a$ is the shift vector, $p_{a b}$ is the conjugate momentum of the induced metric on the preferred foliation, $\hat{k}$ is the trace of the extrinsic curvature of $\mathscr{B}$, and $M_{\textsc{gl}}$ is the mass of the background Lifshitz solution (to be explained in more details below).

To compute the integral~\eqref{eq:HH} for the asymptotically Lifshitz solutions considered in this work, we may begin by choosing the time translation vector along the Killing vector $\chi^a$, for which the shift vector becomes the projection of $\chi^a$ on the leaves of the preferred foliation. Furthermore, by construction, there is no intersection of the `one-boundary' $\mathscr{B}$ with any part of $\dl V$ which itself is a leaf of the preferred foliation (since $n^a = u^a$ on such surfaces). In particular, this kills any possible contribution from the boundaries at past and future infinity. For all possible `inner boundaries' including the universal horizon~\eqref{def:UH}, which are all characterized by the condition $(u\cdot\chi) = 0$ as previously discussed, the vanishing of the integrand in~\eqref{eq:HH} can, in fact, be seen explicitly as follows: for the present choice of the time translation vector, a straightforward computation yields $N = -(u\cdot\chi)$~\cite{Bhattacharyya:2015gwa} and $\hat{k} = \vDl_a s^a \propto (u\cdot\chi)$ and hence $N \hat{k} = 0$ on any $(u\cdot\chi) = 0$ hypersurface; in particular the vanishing of $\hat{k}$ can be appreciated from the fact that $s^a \propto \chi^a$ on any $(u\cdot\chi) = 0$ hypersurface, so that by Killing's equation $\vDl_a s^a$ vanishes here. The second term in the integrand $N^a p_{ab} n^b$ vanishes simply because $p_{a b}$ is a linear combination of the induced metric and the extrinsic curvature of the leaves of the preferred foliation while $n^a = u^a$. Therefore, any contribution to~\eqref{eq:HH} comes~\emph{only} from the part of $\mathscr{B}$ which `resides within' the boundary at infinity $\mathscr{I}$, and this is given by the line generated by the intersection of any preferred slice with $\mathscr{I}$. Moreover, due to the asymptotic alignment of $u^a$ with $\chi^a$, the term containing the shift drops out, so that~\eqref{eq:HH} for our solutions reduces to
	\beqn{
	M = -\frac{1}{8\pi G_{\ae}}\lim_{r\to\infty}\int\limits_{-\infty}^{\infty}\dy\,\frac{r}{\elL}N\hat{k} - M_{\textsc{gl}}~.
	}
As mentioned previously, $M_{\textsc{gl}}$ is the mass of the background Lifshitz solution, and its relevance can be explained as follows: the above expression~\emph{without} the $M_{\textsc{gl}}$ piece, when evaluated on the~\emph{globally} Lifshitz solutions, yields an infinity, whose origin is ultimately the omnipresent vacuum energy. The quantity $M_{\textsc{gl}}$ is precisely this `infinite mass' ascribable to the globally Lifshitz background that needs to be subtracted to make the above expression, applied to an~\emph{asymptotically} Lifshitz solution, meaningful.

The total mass $M$ as given above is still infinite, even with background subtraction, since $\mathscr{B}$ is a non-compact infinite line. As a remedy, we need to regulate the above expression and work with a~\emph{mass per unit length} $M$ of the black hole solutions. To that end, we may modify the above expression as
	\beqn{
	M = -\frac{1}{8\pi G_{\ae}}\lim_{r\to\infty}\lim_{L \to \infty}\frac{1}{L}\int\limits_{-L/2}^{L/2}\dy \frac{r}{\elL}N\hat{k} - M_{\textsc{gl}}~.
	}
where $L$ is a `regulating length' and all the $M$'s now stand for mass per unit length. The integral over the transverse space is now trivial and allows us to cancel out the appearance of the regulating factor $L$. Using the appropriate asymptotic expressions from~\eqref{series:E0-U0-F0}, we then end up with
	\beqn{
	\begin{split}
	M = & -\lim_{r\to\infty}\frac{(r/\elL)^{(z + 1)}}{8\pi G_{\ae}} \times \\
	    & [\frac{1}{\elL} + \frac{(2u_1 - f_1)(z + 1)\elL^{\nu_{\star}-1}}{r^{\nu_{\star}}} + \ord(r^{-2\nu_{\star}})] - M_{\textsc{gl}}~.
	\end{split} 
	}
The leading term proportional to $\elL^{-1}$ is the source of the divergent part in the mass due to the non-zero vacuum energy density as just noted, and $M_{\textsc{gl}}$ is chosen to precisely cancel this term. Once this is taken care of, the spacetime mass per unit length is simply given by
	\beqn{
	M = -\lim_{r\to\infty}\frac{(r/\elL)^{(z + 1)}}{8\pi G_{\ae}} \frac{(2u_1 - f_1)(z + 1)\elL^{\nu_{\star}-1}}{r^{\nu_{\star}}}.
	}
This above quantity goes as $r^{(z + 1) - \nu_{\star}}$. Therefore in the limit $r \to \infty$, the expression for the mass per unit length diverges for $\nu_{\star} < (z + 1)$ while it goes to zero for $\nu_{\star} > (z + 1)$ but is finite and non-zero (in general)~\emph{only} for the choice of
	\beql{def:nu_star}
	\nu_{\star} = (z + 1)~.
	\eeq
We therefore fix $\nu_{\star}$ by requiring that the class of solutions we are studying admits a well-defined, non-zero notion of mass after the appropriate background subtraction. The expression~\eqref{def:nu_star} also is consistent with the standard results for Lifshitz black branes (see \eg~\cite{Taylor:2008tg}). Once this value of $\nu_{\star}$ is used, the mass per unit length of an asymptotically Lifshitz solution is given by
	\beql{def:M:gen}
	M = \frac{(z + 1)(-2u_1 + f_1)}{8\pi G_{\ae}\elL},
	\eeq
where we have expressed everything in terms of canonical quantities $z$ and $\elL$. We will further massage this expression in the next section after we solve the equations of motion for large $r$ and determine the values of $u_1$ and $f_1$.
\subsection{Restrictions on $c_i$ coefficients}
Now that we have fixed $\nu_{\star}$, we may analyze the asymptotic equations of motion. Substituting the expansions~\eqref{series:E0-U0-F0} and~\eqref{series:S0} into the equations of motion~\eqref{eq:poles} and expanding to first order yields a relationship on the $c_i$ coefficients in addition to those established by the global Lifshitz solution~\eqref{global:CC-c14}, namely:
	\beql{eq:c123}
	c_{123} = \frac{4(1 - c_{13})(z - 1)}{n_s(n_s - 2)(z + 1)^2}~.
	\eeq
Therefore, physically acceptable solutions~\emph{only} exist for
	\beql{def:n_s}
	n_s \geqq 3~
	\eeq
and, quite remarkably, asymptotically Lifshitz solutions exist only for `discrete' choices for $c_{123}$. The squared spin-$0$ speed \eqref{speed:s0} then also becomes `quantized' according to
	\beql{q:spin-0:speed}
	s_0^2 = \frac{4z}{(1 - c_{13})[n_s(z + 1) - 2(z - 1)](n_s(z + 1) - 4)}~.
	\eeq
One can then easily show that for $n_s \geqq 3$~\eqref{def:n_s} and for all $z > 1$, the above expression for $s_0^2$ is strictly positive. Hence, all these backgrounds are physically acceptable.

The restriction~\eqref{def:n_s} implies that the analysis of the equations of motion for the next two subleading orders are completely universal. In particular, at $\ord(r^{-(z + 1)})$, we find
	\beql{soln:asymp:o1}
	u_1 = -\frac{\rS}{2\elL}~, \qquad f_1 = 0~, \qquad s_1 = 0~.
	\eeq
Note that the coefficient $u_1$ is actually left undermined, allowing us to trade it for a length scale $\rS$ analogous to the `Schwarzschild radius'. If we plug these values in~\eqref{def:M:gen}, the mass per unit length of the solutions take a cleaner form
	\beql{def:M:can}
	M = \frac{(z + 1)\rS}{8\pi G_{\ae}\elL^2}~.
	\eeq
Next, at $\ord(r^{-2(z + 1)})$ we obtain
	\beql{soln:asymp:o2}
	\begin{split}
	u_2 & = -\frac{\rS^2}{8\elL^2}~, \\
	f_2 & = \frac{(z^2 - 1)\rS^2}{8z^2\elL^2}~, \\
	s_2 & = -\frac{c_{\ae}(n_s - 2)(z + 1)(n_s(z + 1)- 2(z - 1))\rS^2}{32(n_s + 1)z^2\elL^2}~.
	\end{split}
	\eeq
The analysis, for a completely general $n_s$, can only be carried out until this order, as already observed; to proceed further one needs to pick an $n_s$. However, the general feature of all such solutions are similar: the solution will initially depend on two free parameters, namely $\rS$ and $c_{\ae}$. We have already noted that $\rS$ is directly related to the mass of the solution~\eqref{def:M:can}. Although one cannot do this analytically, in principle $c_{\ae}$ is fixed by demanding regularity on the spin-$0$ horizon, \ie setting $F_X|_{s_0} = F_e|_{s_0} = 0$. This leaves the a one parameter family of solutions specified by $\rS$. Furthermore, this analysis makes clear that we must choose particular values for the $c_i$ coefficients in our numerical evolution or we will not asymptote to a Lifshitz solution. We now turn to the numerical procedure.
\section{Asymptotically Lifshitz black holes}\label{LifBH}
The equations~\eqref{eq:poles} do not yield exact solutions with Lifshitz asymptotics and universal horizons, so we have to resort to numerics. The basic approach is described in details below below and closely follows the route taken in~\cite{Barausse:2011pu}.

To begin with, we need to make choices for the various couplings and paramaters. In this work, we have only focussed on the case of $z = 2$, although asymptotically Lifshitz solutions with $z > 2$ are expected to have qualitatively similar features. We will also set $\elL = 1$ without any loss in generality. Finally, we will apply a field redefinition and choose the values of $c_i$ so that the spin-$0$ and Killing horizons are colocated. Since the Lifshitz exponent $z$ is determined solely by $c_{14}$~\eqref{global:CC-c14}, the disformal field redefinitions~\eqref{eq:coeffdisformal}, which do not change $c_{14}$, preserve the Lifshitz exponent $z$ as well, and hence also preserves $\elL$~\eqref{global:CC-c14}. However, as mentioned previously, the location of the spin-$0$ horizon can be shifted. We use the field redefinition to colocate the spin-$0$ and Killing horizon, and then choose coefficients that satisfy the `discreteness' condition on $c_{123}$~\eqref{eq:c123}. Our numerical results in this section and section~\ref{Smarr} are for the coefficients choice
	\beql{eq:coeffs}
	c_{14} = \frac{1}{2}~, \qquad c_{13} = \frac{9}{10}~, \qquad c_2 = -\frac{161}{180}~,
	\eeq
and for $n_s = 4$. One may check that for the above choice of coeffients one gets $s_0 = 1$ from~\eqref{q:spin-0:speed}.

With the above choices, our approach for finding asymptotically Lifshitz solutions is as follows:
	\begin{enumerate}
	\item Analytically expand the equations of motion about the spin-$0$ horizon and solve for $e(r)$ and $e'(r)$ in terms of $X(r)$ and $X'(r)$ there so that $e''(r)$ and $X''(r)$ remain regular.
	\item Evolve outwards and inwards from the spin-$0$ horizon numerically.
	\item Iterate (\emph{\`a la} the `shooting method') $e'(r)$ and $X'(r)$ while keeping $e(r)$ and $X(r)$ fixed at the spin-$0$ horizon until the solution is asymptotically Lifshitz.
	\item Perform an overall normalization on the solution, which corresponds to choosing and initial value of $X(r)$ on the spin-$0$ horizon so that $r^{-4}e(r) \to 1$ and $r^{-2}X(r) \to 1$ as $r \to \infty$.
	\end{enumerate}
We now address each of these steps and then present some example numerical results.
\subsection{Analytic near spin-$0$/Killing horizon expansion}
With the above choice of coefficients the singularity in the equations of motion occur at the Killing horizon since the spin-$0$ horizon is colocated. The spin-$0$ horizon location $r_{s_0}$ is a free parameter at this point but will eventually be related to the mass of the spacetime. At $r_{s_0}$, the value of $X(r)$ can also be chosen freely as it just changes the overall scale of the eventual solution, but $e(r_{s_0}) = 0$ by definition of a Killing horizon (recall $e(r) = -\met_{a b}\chi^a\chi^b$). We then analytically expand $e(r)$ and $X(r)$ as a power series in $(r - r_{s_0})$ out to fourth order. Solving the equations of motion analytically order by order and imposing regularity by requiring that $F_e$ and $F_X$ vanish on the spin-$0$ horizon relates the coefficients for the near horizon expansion of $e(r)$, $e'(r)$, $X(r)$, and $X'(r)$. All the coefficients are fixed other than a dependence on a single additional, undetermined parameter $\mu$, which exists in addition to $r_{s_0}$ since we have at this point only imposed regularity at the spin-$0$ horizon but haven't specified the asymptotic behaviour of the solutions. As mentioned previously, only by requiring Lifshitz asymptotics \textit{and} spin-$0$ horizon regularity are we able to reduce the solutions to a one parameter family. We start the evolution at $r = r_{s_0}(1 \pm \ord(10^{-5}))$ which yields an initial accuracy in $e(r)$, $e'(r)$, $X(r)$ and $X'(r)$ vs. the exact solution of $\ord(10^{-20})$.
\subsection{Numerical evolution and normalization}
Given the initial values $X(r_{s_0})$, $r_{s_0}$, and $\mu$ we evolve outwards from $r = r_{s_0}(1 \pm \ord(10^{-5}))$ respectively with Mathematica. For generic values of $\mu$ the exterior solution eventually significantly deviates from the Lifshitz geometry and, in fact, breaks down at some radius $r_{dev}$. We search in the $\mu$ parameter space, which changes $e'(r_{s_0})$ and $X'(r_{s_0})$, to maximize $r_{dev}$. In principle, by tuning $\mu$ arbitrarily finely we can push $r_{dev}$ out to infinity and land on the `exact' asymptotically Lifshitz solution. In practice we tune $\mu$ until $r_{dev}$ is at least a factor of $10^4$ larger than $r_{s_0}$. This gives a very accurate asymptotic Lifshitz region. It also, as promised, reduces the solution space to a one parameter family controlled by $r_{s_0}$. Evolving inwards with this $\mu$ then determines $\ruh$.

The initial value of $X(r)$ controls the overall scaling of $e(r)$, and $X(r)$ in the solution. After each solution with some initial $r_{s_0}$ has been found, we scale the asymptotic solution such that the leading order term in $e(r)$ goes exactly as $r^4$.
\subsection{Example solution}
\begin{figure}[htb]
	 \centering
	 \includegraphics[scale=0.58]{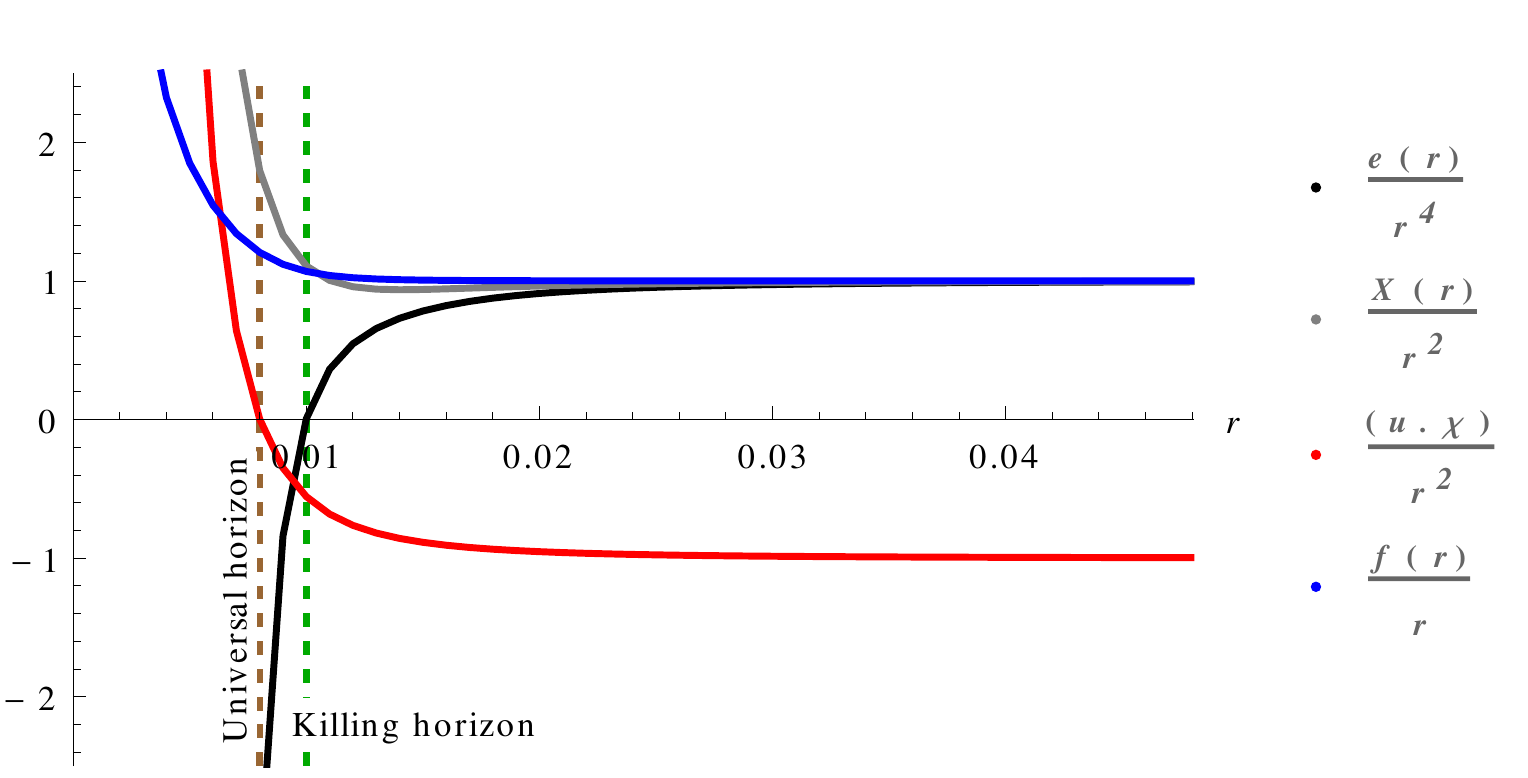}
	 \caption{Normalized metric and {\ae}ther coefficients for an asymptotically Lifshitz solution. Here $c_{14} = \frac{1}{2}$, $c_{13} = \frac{9}{10}$, $c_2 = -\frac{161}{180}$, units are chosen such that $\elL = 1$, and the radius of the spin-$0$/Killing horizon is $r_{s_0} = 0.01$. The radial location of the universal horizon is $\ruh \approx 0.008$ in these units.}
	 \label{fig:examplesolution}
	\end{figure}

At the end of our procedure we have a full solution over the entire spacetime for the functions $e(r)$ and $X(r)$ (and hence all other functions with can be expressed in terms of $e(r)$, $X(r)$ and their derivatives) that is asymptotically Lifshitz and possesses a universal horizon. The Lifshitz normalized coefficients (\ie dividing the coefficients by their approriate scaling in the globally Lifshitz case) for a typical solution is given in Fig.~\ref{fig:examplesolution}.
Note that numerical evolution inside the universal horizon is possible in this construction and indeed Fig.~\ref{fig:examplesolution} shows the behavior of the free metric and {\ae}ther components inside but still near the universal horizon. In principle, solutions can admit multiple $(u\cdot\chi) = 0$ hypersurfaces (c.f.~\cite{Barausse:2011pu}). In such a case, the outermost $(u\cdot\chi) = 0$ hypersurface denotes the universal horizon, as that is the surface that causally separates asymptotic infinity from an interior region. Since we are interested solely in the behaviour of `outside region' of the spacetime up to the universal horizon, we have not categorized the interior structure of our solutions in detail. 
\section{Mass and the first law}\label{Smarr}
For each numeric solution we fit the numerical solutions for $e(r)$ and $X(r)$ by their asymptotic expansions in section~\ref{sec:asymptotics} out to fifth order in $r^{-(z + 1)}$. In particular, this yields the corresponding coefficients $f_1$ and $u_1$ and $s_1$ in the asymptotic solutions~\eqref{soln:asymp:o1}, and we find $s_1$ and $f_1$ to be zero within the desired accuracy (thereby providing a consistency check on the numerical evolutions). The value of $u_1$ provides the value of the dimensionful parameter $\rS$ hence allowing us to compute the mass per unit length from~\eqref{def:M:can} for each one of the numerically constructed solutions.

\begin{figure}[htb]
	 \centering
	 \includegraphics[scale=0.7]{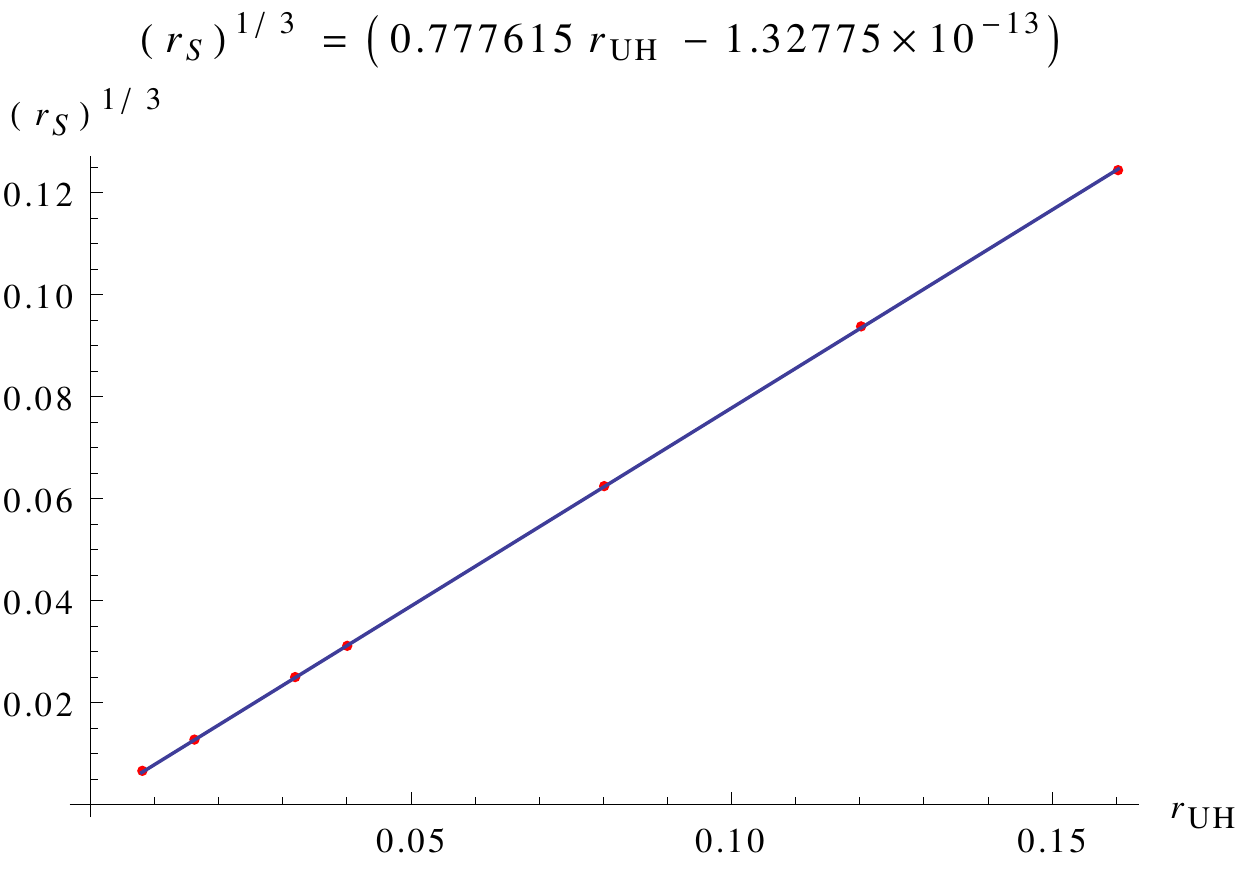}
	 \caption{$\rS^{1/3}$ vs. $\ruh$ for $z = 2$ asymptotically Lifshitz black holes with a universal horizon.}
	 \label{fig:r0vsruh}
	\end{figure}

We evaluate how the mass scales with the radius of the universal horizon by calculating multiple numerical solutions with different initial values of the spin-$0$/Killing horizon and fitting the resulting $\rS$ and $\ruh$ values. In figure~\ref{fig:r0vsruh} we can see that up to a tiny numerical error, $\rS$ and hence the mass per unit length $M$, homogeneously scales as $\ruh^3$.
Since the mass is a homogeneous function of $\ruh$ a first law of the form $\delta M = q\delta A$, where $\delta A = \delta \ruh$ and $q \propto \ruh^2$ is guaranteed. Note that the homogeneity of $M$ with respect to $\ruh$ is a non-trivial result as there is an extra scale, the Lifshitz scale, involved and therefore there is no guarantee of homogeneity~\emph{a priori}. Indeed, failure of homogeneity occurs in asymptotically AdS solutions in $D = 1 + 3$~\cite{Bhattacharyya:2014kta}. In the $D = 1 + 3$ case, the first law for the asymptotically AdS solutions is of the form $\delta M = q \delta A$, but $q$ is a non-homogeneous and indeed non-analytic function of $\ruh$ (see equation (58) of~\cite{Bhattacharyya:2014kta}) without any obvious thermodynamic interpretation. Therefore while one might have expected that a similar failure of na\"ive thermodynamics happens in the Lifshitz case as well, since AdS can be thought of simply as a $z = 1$ Lifshitz spacetime, this turns out to be incorrect. Rather, as we shall see below, the Lifshitz solutions hold the possibility of a much more natural thermodynamic interpretation.

If a thermodynamic interpretation of the first law for the above Lifshitz solutions exists, the temperature of the universal horizon must scale as $\ruh^2$. Previous work on static, spherically symmetric universal horizon solutions with flat asymptotics indicated that the temperature of the universal horizon calculated locally using the tunneling approach~\cite{Berglund:2012fk} is given by $T = (a\cdot\chi)_{\textsc{uh}}/4\pi$. By considering the peeling of non-relativistic high energy modes (those with very high group velocity) near the universal horizon~\cite{Cropp:2013sea}, one can define an appropriate notion of surface gravity, $\kappa_{\textsc{uh}} = (a\cdot\chi)_{\textsc{uh}}/2$, which yields the familiar relationship $T = \kappa_{\textsc{uh}}/2\pi$. Since both constructions are local, one would expect that they are independent of asymptotics and the temperature in asymptotically Lifshitz solutions is also proportional to $(a\cdot \chi)_{\textsc{uh}}$. For the present case, therefore we need $(a\cdot\chi)_{\textsc{uh}} \propto \ruh^2$ as only with this scaling is it possible to construct a first law of the standard form.

\begin{figure}[htb]
		\centering
		\includegraphics[scale=0.7]{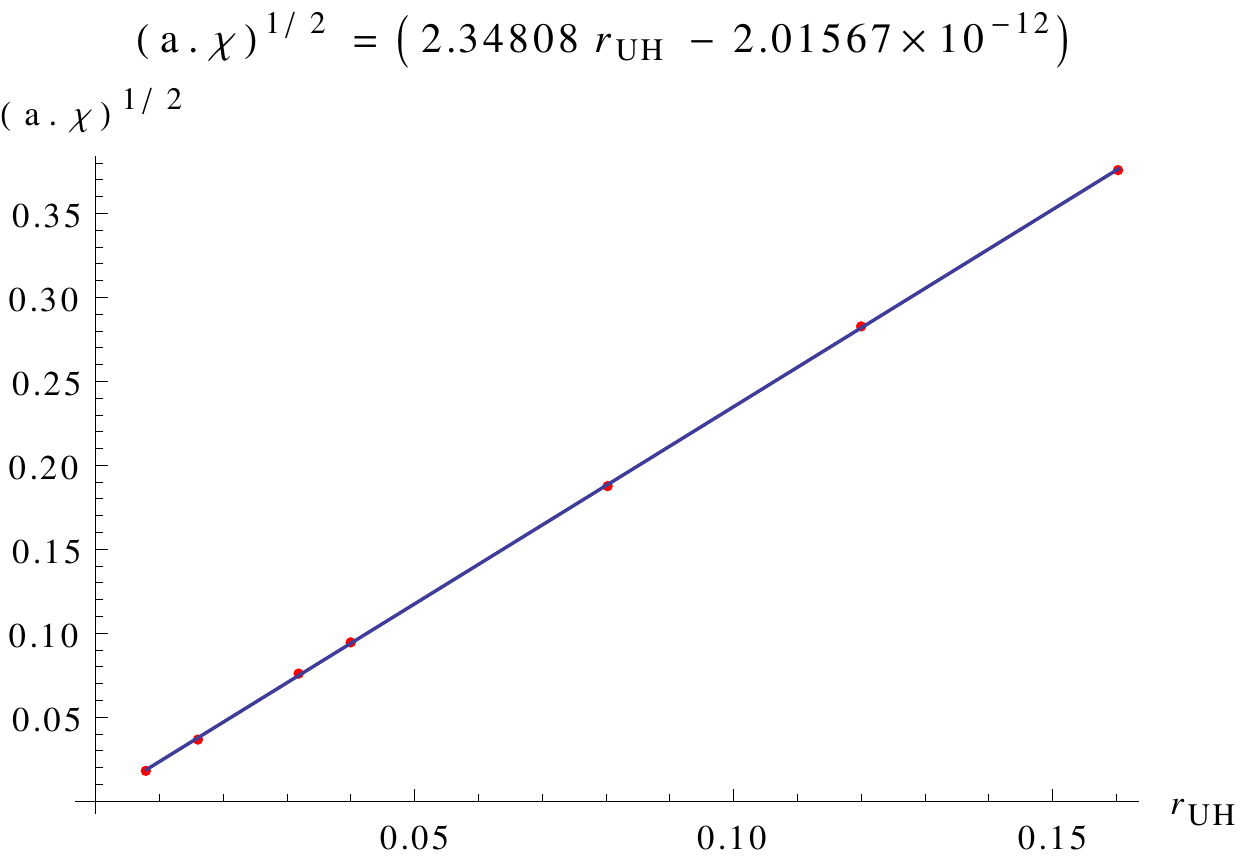}
		\caption{$(a\cdot\chi)_{\textsc{uh}}^{1/2}$ vs. $\ruh$ for $z = 2$ asymptotically Lifshitz black holes with a universal horizon.}
		\label{fig:adotchivsruh}
	\end{figure}

We show in figure~\ref{fig:adotchivsruh} that $(a\cdot\chi)_{\textsc{uh}}$ has precisely the correct scaling with $\ruh^2$ to construct the first law. Therefore the first law for asymptotically Lifshitz solutions is at least compatible with a straightforward thermodynamic interpretation. A full verification of thermodynamics for Lifshitz solutions would, of course, require a calculation of the temperature in this case as well. For now we merely state that the first law of mechanics is compatible with such a thermodynamic interpretation and that all indications are that a first law of the form $\delta M = T \delta S$, with $T \propto (a\cdot\chi)_{\textsc{uh}}$, holds. We stress again that this is very different from the asymptotically AdS case. The origin of this discrepancy remains, at the moment, a mystery.
\section{Conclusion}\label{Conclusions}
We have analyzed and constructed a new class of solutions in $D = (1 + 2)$ dimensional Ho\v{r}ava gravity and Einstein-{\ae}ther theory, those with universal horizons and Lifshitz asymptotics. For at least $z = 2$ asymptotics there is a first law of mechanics that fits nicely with what is known about universal horizon thermodynamics. This is in contrast to the $D = 1 + 3$ asymptotically AdS case, where the first law does not have a straightforward thermodynamic interpretation. Of course, one still needs to calculate a temperature for Lifshitz solutions to complete a thermodynamical relationship, which we leave for future work. If such a thermodynamics holds, these solutions would then provide an interesting playground for explorations of Lifshitz holography. The structure of both the asymptotic and near horizon regions is dramatically different from what is found in the usual AdS$_3$/CFT$_2$ construction -- neither region has a symmetry algebra appropriate to a (relativistic) conformal field theory. Therefore neither the state counting approaches used at the boundary in $D = 1 + 2$ gravity for BTZ black holes or near Killing horizons in higher dimension, which rely on establishing invariance under a Virasoro algebra, na\"ively apply. We shall return to this question of calculating the entropy of a universal horizon using Lifshitz algebras in future work.\\
\\
\begin{acknowledgments}
JB acknowledges support from Prof. Thomas Sotiriou, University of Nottingham, UK via funding from the European Research Council under the European Union’s Seventh Framework Programme (FP7/2007-2013) / ERC grant agreement n. 306425 “Challenging General Relativity”.
\end{acknowledgments}

\end{document}